\documentclass[aps,prl,twocolumn,groupedaddress,showpacs]{revtex4}% PRL
\usepackage{amsfonts}
\usepackage{mathrsfs}
\usepackage{graphicx}
\usepackage{subfigure}
\usepackage{amsmath}
\usepackage{amssymb}
\usepackage{amsbsy}
\usepackage{bm}

\def\bs{\bm{\sigma}}
\def\bx{\bm{\xi}}

\def\hm{\hat{m}}

\begin{document}

\title{Role of zero synapses in unsupervised feature learning}

\author{Haiping Huang}
%\email{physhuang@gmail.com}
\affiliation{RIKEN Brain Science Institute, Wako-shi, Saitama
351-0198, Japan}
\date{\today}

\begin{abstract}
Synapses in real neural circuits can take discrete values including zero (silent or potential) synapses. The computational role of zero
synapses in unsupervised feature learning of unlabeled noisy data is still unclear, thus it is important to understand how the sparseness of synaptic activity is shaped
during learning and its relationship with receptive field formation. Here, we formulate this kind of sparse feature learning by a
statistical mechanics approach. We find that learning decreases the fraction of zero synapses, and when the fraction decreases rapidly around
a critical data size, an intrinsically structured receptive field starts to develop. Further increasing the data size refines the receptive field, while 
a very small fraction of zero synapses remain to act as contour detectors. This phenomenon is discovered not only in learning a handwritten digits dataset, but also in learning retinal
neural activity measured in a natural-movie-stimuli experiment. 
\end{abstract}

 \maketitle

%%%%%%%%%%%%%%%%%%%%%%%%%%%%%%%%%%%%%%%%%%%%%%%%%%%%%%%%%%%%%%%%%
\section{Introduction}
Sparsity in either neural activity or synaptic connectivity plays an important role in sensory information processing across brain areas~\cite{Len-2003,Huang-2016a}.
Sparsity constraints imposed on neural activity in a sparse coding model~\cite{SC-1996} reproduce Gabor-like filters (edge detectors), which resemble receptive fields of
simple cells in the mammalian primary visual cortex. The sparse representation was also applied to deep belief networks to model hierarchical representations of natural image
statistics~\cite{Lee-2008}, which capture higher order features at deeper levels of the cortical hierarchy. In addition, from a perspective of optimal information
storage, there must exist a large fraction of silent or potential synapses~\cite{Brunel-2004}, consistent with the existence of these synapses in cortex and cerebellum~\cite{Barb-2007}.
These (zero) synapses are vital for plasticity during learning~\cite{Brunel-2004}. Therefore, the sparse representation in synaptic connectivity is also appealing in
optimal neural computation.

Most artificial neural networks are trained by supervised learning, which requires a large library of images prelabeled with categories. However, 
unsupervised learning gives humans and non-human animals the ability to make sense of the external world by themselves, without any additional supervision. Thus unsupervised learning aims at extracting regularities
in sensory inputs without specific labels. To figure out what learning
algorithms may be used for modeling the external world in an unsupervised way is important for designing machine intelligence. However, understanding computational mechanisms of unsupervised
learning in sensory representation is extremely challenging~\cite{Erhan-2010}.

Although zero synapses were observed in real neural circuits, their computational role in concept formation during unsupervised learning remains unclear,
and lacks a simple model to explain. Furthermore,
previous theoretical efforts focused on random models of neural networks~\cite{Barra-2012,Barra-2014,Remi-2016}, where the distribution of synaptic values is prefixed.
Here, we propose a simple model of unsupervised learning with zero synapses, where a two-layered neural
network is introduced to learn the synaptic values from  sensory inputs, which is thus more practical than random models. The bottom layer is composed of visible neurons receiving sensory inputs, 
while the top layer contains only one hidden neuron in response to specific features in the inputs. The bottom layer is connected to the top layer by synapses. Note that there do not exist lateral connections in the bottom layer. This kind of neural network is called a one-bit restricted Boltzmann machine (RBM)~\cite{Smolensky-1986,Huang-2016b}.
Binary synapses were also experimentally observed in real neural circuits~\cite{Peter-1998,Connor-2005}. The one-bit RBM with binary synapses has been studied as a toy model of unsupervised 
feature learning~\cite{Huang-2016b}, and is analytically tractable at the mean-field level~\cite{Huang-2017}.

To study the computational role of zero synapses, we model the connections between bottom and top layers in the one-bit RBM by ternary synapses, 
which take discrete values of $\{0,\pm1\}$. Given a sensory input, 
the ternary synaptic connections provide a hidden feature representation of the input. One configuration of ternary synapses forms a
feature map and is also called the receptive field of the hidden neuron at the top layer.

\section{Problem setting and mean-field method}
In this study, we use the one-bit RBM defined above to learn specific features in sensory inputs, which are raw unlabeled data. The machine is required to internally create concepts
about the inputs. This process is thus called unsupervised learning. Here,
the sensory inputs are given by handwritten digits taken from the MNIST dataset~\cite{Lecun-1998}. Each image from this dataset has $28\times28$ pixels, specified by an Ising-like spin configuration $\bs=\{\sigma_i=\pm1\}_{i=1}^{N}$ where $N$ is the input dimensionality.
A collection of $M$ images is denoted by $\{\bs^a\}_{a=1}^{M}$. The number of synapses is the same as the input dimensionality. Synaptic values are characterized by $\bx$, where each component takes one of the ternary values. The one-bit RBM is thus described by the Boltzmann distribution $P(\bs,s)\propto\exp\left[\frac{\beta}{\sqrt{N}}\sum_i\xi_i\sigma_is\right]$,
where $s=\pm1$ denotes the activity of the hidden neuron, $\beta$ denotes an inverse temperature, and the synaptic strength is scaled by the factor $\frac{1}{\sqrt{N}}$ to
ensure that the corresponding statistical mechanics model has an extensive free energy. 
After marginalization of the hidden activity, one obtains the distribution of the visible activity as:
\begin{equation}\label{Pobs0}
 P(\bs|\bx)=\frac{\cosh\Bigl(\frac{\beta}{\sqrt{N}}\bx^{{\rm T}}\bs\Bigr)}{\sum_{\bs}\cosh\Bigl(\frac{\beta}{\sqrt{N}}\bx^{{\rm T}}\bs\Bigr)}.
\end{equation}

As an inference model, for any given input, the one-bit RBM has $3^N$ possible synaptic configurations to describe the sensory input. However, the machine will choose one of these
potential candidates as the feature map. This process is naturally modeled by Bayes' rule:
\begin{equation}\label{Pobs}
\begin{split}
 &P(\bx|\{\bs^{a}\}_{a=1}^{M})=\frac{\prod_{a}P(\bs^{a}|\bx)}{\sum_{\bx}\prod_{a}P(\bs^{a}|\bx)}\\
 &=\frac{1}{Z}\prod_{a}\cosh\left(\frac{\beta}{\sqrt{N}}\bx^{{\rm T}}\bs^{a}\right)\prod_{i}e^{-M\ln\cosh\Bigl(\frac{\beta\xi_i}{\sqrt{N}}\Bigr)},
\end{split}
 \end{equation}
where $Z$ is the partition function of the model, and
a uniform prior for $\bx$ is assumed. $\beta$ serves as the inverse-temperature like parameter to control learning noise. The synaptic scaling also requires $\beta\ll\sqrt{N}$, and under this condition, the last product in Eq.~(\ref{Pobs})
becomes $\prod_i e^{-\gamma\xi_i^2}$, where $\gamma\equiv\frac{\alpha\beta^2}{2}$ with $\alpha\equiv\frac{M}{N}$. It is clear that from a Bayesian viewpoint, introducing zero synapses amounts to
some sort of Gaussian-like regularization but with discrete support. The Bayesian
method is able to capture uncertainty in learned parameters (synaptic values here) and thus avoids over-fitting~\cite{MacKay-1992}. It may be able to reduce the necessary data size for learning as well~\cite{Huang-2016b}.

In what follows, we compute the maximizer of the posterior marginals estimator $\hat\xi_i = \arg\max_{\xi_i}P_i(\xi_i)$~\cite{Nishimori-2001}, where the 
data dependence of the probability is omitted.
Hence, the task is to compute marginal probabilities, e.g., $P_i(\xi_i)$, which is a computationally hard problem due to the interaction among data constraints (the product over $a$ in Eq.~(\ref{Pobs})). However, by mapping
the original model (Eq.~(\ref{Pobs})) onto a graphical model~\cite{Huang-2016b}, where data constraints and synaptic values are treated as factor (data) nodes and variable nodes respectively, one can estimate the marginal probability by running a message passing algorithm as we shall explain below.
The key assumption is that synapses on the graphical model are weakly correlated, which is called the Bethe approximation~\cite{MM-2009} in physics. We first define a cavity
probability $P_{i\rightarrow a}(\xi_i)$ of $\xi_i$ with data node $a$ removed. Under the weak correlation assumption, $P_{i\rightarrow a}(\xi_i)$
satisfies a self-consistent equation:
\begin{subequations}\label{bp0}
\begin{align}
P_{i\rightarrow a}(\xi_{i})&=\frac{1}{Z_{i\rightarrow a}}
e^{-\gamma\xi_i^2}\prod_{b\in\partial i\backslash
a}\mu_{b\rightarrow i}(\xi_{i}),\label{bp2}\\
\begin{split}
\mu_{b\rightarrow i}(\xi_{i})&=\sum_{\{\xi_{j}|j\in\partial
  b\backslash
  i\}}\cosh\left(\frac{\beta}{\sqrt{N}}\bx^{{\rm T}}\boldsymbol{\sigma}^{b}\right)\prod_{j\in\partial
  b\backslash i}P_{j\rightarrow b}(\xi_{j}),\label{bp1}
\end{split}
\end{align}
\end{subequations}
where $Z_{i\rightarrow a}$ is a normalization constant,
$\partial i\backslash a$ denotes neighbors of feature node $i$ except
data node $a$, $\partial b\backslash i$ denotes neighbors of
data node $b$ except feature node $i$, and the auxiliary quantity
$\mu_{b\rightarrow i}(\xi_i)$ indicates the probability contribution from
data node $b$ to feature node $i$ given the value of
$\xi_i$~\cite{MM-2009}. Products in Eq.~(\ref{bp0}) stem from the weak correlation assumption.

In the thermodynamic limit, the sum inside the hyperbolic cosine function in Eq.~(\ref{bp1}), excluding the $i$-dependent term, is a random variable following a normal distribution with mean $G_{b\rightarrow i}$ and variance
$\Xi_{b\rightarrow i}^{2}$~\cite{Huang-2015b}, where $G_{b\rightarrow
i}=\frac{1}{\sqrt{N}}\sum_{j\in\partial b\backslash i}\sigma_{j}^{b}m_{j\rightarrow b}$ and
$\Xi^{2}_{b\rightarrow i}\simeq\frac{1}{N}\sum_{j\in\partial b\backslash
i}(\hm_{j\rightarrow b}-m_{j\rightarrow b}^{2})$. The cavity magnetization is defined as $m_{j\rightarrow b}=\sum_{\xi_j}\xi_jP_{j\rightarrow b}(\xi_j)$, while the second moment of the feature component $\xi_j$ is defined
by $\hm_{j\rightarrow b}\equiv\sum_{\xi_j}\xi_j^2P_{j\rightarrow b}(\xi_j)$. Thus the intractable sum over
all $\xi_j$ ($j\neq i$) can be replaced by an integral over the normal distribution. Furthermore, because $\xi_i$ is a ternary variable, $P_{i\rightarrow a}(\xi_i)$ 
can be parametrized by cavity fields $h_{i\rightarrow a}$ and $g_{i\rightarrow a}$, as $P_{i\rightarrow a}(\xi_i)=\frac{e^{\xi_ih_{i\rightarrow a}+(\xi_i^2-1)g_{i\rightarrow a}}}
{e^{h_{i\rightarrow a}}+e^{-h_{i\rightarrow a}}+e^{-g_{i\rightarrow a}}}$. Combining this representation with Eq.~(\ref{bp0}), we have the following iterative learning equations:
\begin{subequations}\label{bp3}
\begin{align}
m_{i\rightarrow a}&=\frac{\zeta_{i\rightarrow a}}{1+\zeta_{i\rightarrow a}}\tanh h_{i\rightarrow a},\label{bp3a}\\
h_{i\rightarrow a}&=\sum_{b\in\partial i\backslash a}\tanh^{-1}\Bigl(\tanh\beta G_{b\rightarrow i}\tanh\frac{\beta}{\sqrt{N}}\sigma_i^b\Bigr),\\
\zeta_{i\rightarrow a}&=e^{-\gamma}\sum_{x=\pm1}\prod_{b\in\partial i\backslash a}e^{u^x_{b\rightarrow i}},\label{bp3c}
\end{align}
\end{subequations}
where $u^x_{b\rightarrow i}\equiv\ln\Bigl[\cosh\frac{\beta\sigma_i^b}{\sqrt{N}}(1+x\tanh\beta G_{b\rightarrow i}\tanh\frac{\beta\sigma_i^b}{\sqrt{N}})\Bigr]$, and $2\cosh h_{i\rightarrow a}e^{g_{i\rightarrow a}}=\zeta_{i\rightarrow a}$.
$m_{i\rightarrow a}$ can be interpreted as the message passing from feature node $i$ to data node $a$, while $u^x_{b\rightarrow i}$ can be
interpreted as the message passing from data node $b$ to feature node $i$, depending on $x$. Note that the prefactor $\frac{\zeta_{i\rightarrow a}}{1+\zeta_{i\rightarrow a}}$ 
in Eq.~(\ref{bp3a}) is the cavity probability of non-zero synapses, i.e., $1-P_{i\rightarrow a}(0)$, and this is also $\hm_{i\rightarrow a}$ according to the definition. In this sense, the sparsity of synapses is described by
a single parameter $\rho\equiv\frac{1}{N}\sum_iP_i(0)$, where $P_i(0)=\frac{1}{1+\zeta_i}$. The potential feature (synaptic configuration) is inferred by computing 
$\zeta_i=e^{-\gamma}\sum_{x=\pm1}\prod_{b\in\partial i}e^{u^x_{b\rightarrow i}}$ as well as
$m_i=\frac{\zeta_i}{1+\zeta_i}\tanh\left(\sum_{b\in\partial i}\tanh^{-1}\Bigl(\tanh\beta G_{b\rightarrow i}\tanh\frac{\beta}{\sqrt{N}}\sigma_i^b\Bigr)\right)$, in which the magnetization $m_i$ is related to $P_i(1)$ via $P_i(1)=\frac{1+m_i-P_i(0)}{2}$.

If the weak correlation assumption is
self-consistent, starting from randomly initialized messages, the learning equations will converge
to a fixed point corresponding to a thermodynamically dominant minimum of the Bethe free energy function ($-\frac{1}{\beta}\ln Z$)~\cite{MM-2009}.
In the following part, we study how the learned feature map and the fraction of zero synapses change with data size. In particular, we focus on when
the machine develops an internal concept about the input handwritten digits and what the computational role of zero synapses is for feature selectivity and receptive field
formation.
%%%%%%%%%%%%%%%%%%%%%%%%%%%%%%%%%%%%%%%%%%%%%%%%%%%%%%%%%%%%%%%%%%%%
\begin{figure}
\centering
 (a) \includegraphics[bb=0 0 428 308,scale=0.6]{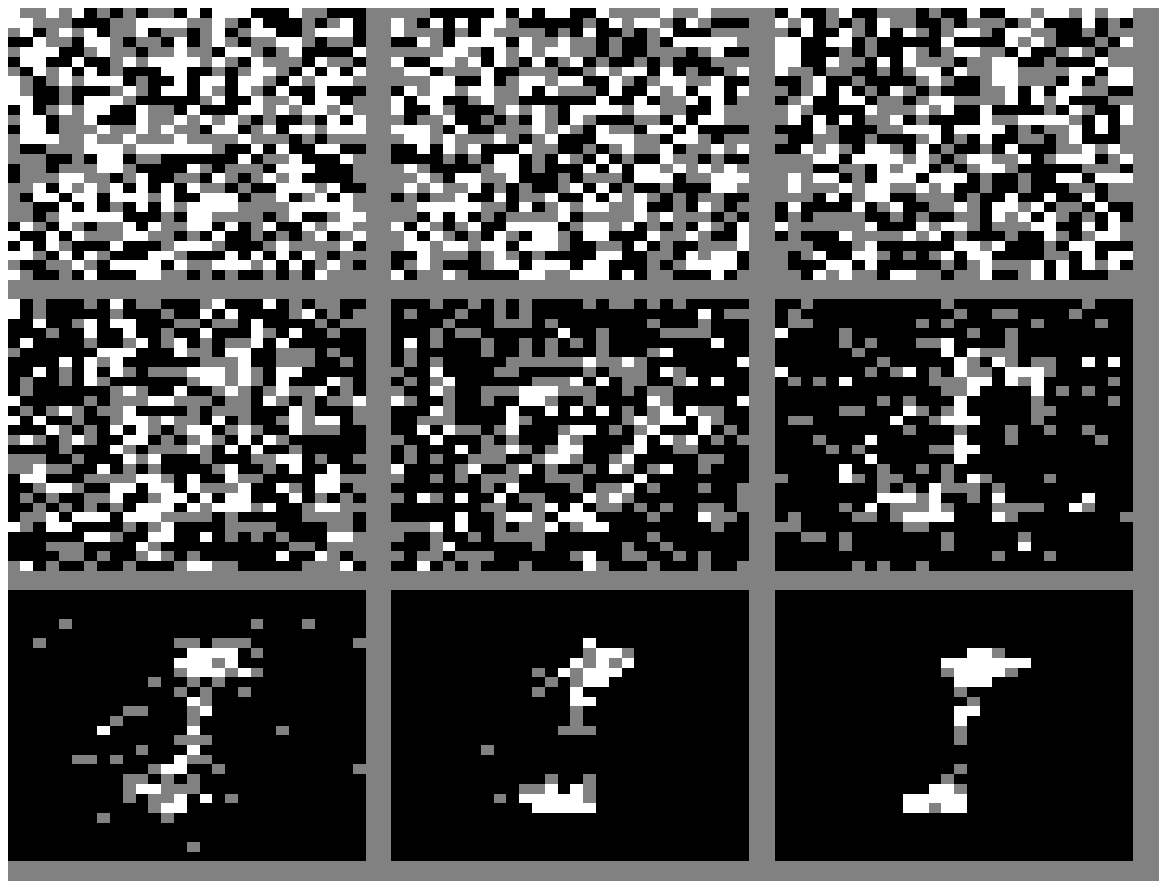}
     \hskip .05cm
     \includegraphics[bb=0 0 325 244,scale=0.7]{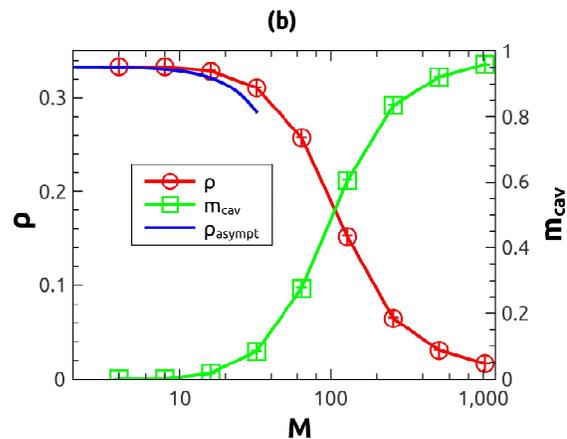}
  \caption{(Color online) Learning behavior in a handwritten digits dataset (digits $0$ and $1$). (a) Formation of receptive fields during learning. From left to right and top to
  bottom, the corresponding data size increases as $M=2^k$ ($k=2,3,\ldots,10$). The colors black, white and gray indicate inactive ($\xi_i=-1$), active ($\xi_i=1$), and zero ($\xi_i=0$) synapses respectively.
  The gray border separating subfigures does not refer to synapses.
  (b) The fraction $\rho$ of zero synapses (left axis) and the overall strength of cavity messages ($m_{{\rm cav}}$, right axis) as a function of data size. Each marker in the plot is averaged over ten random selections of training images with equal number.
  The asymptotic curve of $\rho$ in the small limit of $m_{{\rm cav}}$ is also shown.
      }\label{RFs01}
 \end{figure}

\section{Results}
%%%%%%%%%%%%%%%%%%%%%%%%%%%%%%%%%%%%%%%%%%%%%%%%%%%%
We use the above mean field theory to study unsupervised feature learning with zero synapses. In the following simulations, $\beta=0.5$ unless otherwise stated. For simplicity, we consider only the  $0$ and $1$ digits, because other combinations of two different digits yield similar results. We first study
how the receptive field of the hidden neuron develops during the learning, as the number of training images increases. As shown in Fig.~\ref{RFs01} (a), when the data is severely
scarce, there is no apparent structure in the feature map. When the number of training images increases up to around $100$, an intrinsically structured feature map starts to develop. Nevertheless,
there are still a large fraction of zero synapses. As learning proceeds, the intrinsic structure concentrates more on the center of the feature map, indicating that the
machine has already created an internal perception of external stimuli. This kind of perception has been shown to have an excellent discriminative 
capability on different stimuli by a precision-recall analysis~\cite{Huang-2016b}. This is because the distribution of the weighted sum of inputs the hidden neuron receives develops two well-separated peaks for two 
different digits.

Then, we study the computational role of zero synapses. As shown in Fig.~\ref{RFs01} (b), the sparsity level of synapses $\rho$ decreases with the data size $M$. 
Around some critical $M$, the sparsity decreases abruptly, suggesting that a structured feature map is beginning to develop. Here, learning indeed induces the fraction of zero synapses to
decrease~\cite{Brunel-2004}, since some zero synapses are required to adopt non-zero values for capturing characteristics in the input data. When the data size is further increased,
the feature map is refined, and the sparsity decreases more slowly than around the critical region. At large values of $M$, a small fraction of zero synapses are still
maintained. The zero synapses at this stage seem to form an approximate boundary between active and inactive regions in the feature map (see the last
feature map in Fig.~\ref{RFs01} (a)). Therefore, the zero synapses behave like contour detectors. This effect is predicted by our model, but its neurobiological counterpart is still unclear and deserves tested in future experiments of feature learning.

In particular, the monotonic behavior of the sparsity level of synapses is intimately related to the overall strength of cavity messages, which is defined as $m_{{\rm cav}}\equiv\frac{1}{MN}\sum_{(i,a)}\tanh^2h_{i\rightarrow a}$. The model has originally
symmetry, since Eq.~(\ref{Pobs}) is invariant under the transformation of $\bx\rightarrow-\bx$. This symmetry can be spontaneously broken, as indicated by $m_{{\rm cav}}$ (Fig.~\ref{RFs01} (b)).
Around some critical $M$, cavity messages start to polarize without maintaining trivial (null) values any more, which is accompanied by the rapid decrease of the number of zero synapses.
Moreover, the asymptotic behavior of $\rho$ in the small limit of
the message strength can be derived as $\rho_{{\rm asympt}}=\frac{1}{3}[1-\frac{\beta^4\epsilon^2\omega}{3}M^2]$, where $\epsilon$ denotes the small strength ($m_{j\rightarrow b}=\epsilon$, $\forall(j,b)$), and $\omega$ denotes the image statistics expressed as $\omega=\frac{1}{N}\sum_if_i^2$ in which
$f_i=\frac{1}{MN}\sum_{b,j}\sigma_i^b\sigma_j^b$. This asymptotic behavior captures well the trend of $\rho$ when $M$ is small (Fig.~\ref{RFs01} (b)).

\begin{figure}
\centering
     \includegraphics[bb=0 0 428 308,scale=0.57]{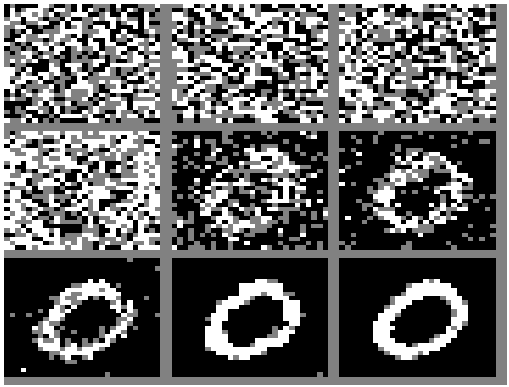}
  \caption{
  (Color online) Receptive field formation when only images of various digit $0$ are learned without any label information. From left to right and top to
  bottom, the corresponding data size increases as $M=2^k$ ($k=2,3,\ldots,10$).
  }\label{RFs0}
\end{figure}

Next, we study how the machine creates a perception of only one digit such as $0$, as learning proceeds. Receptive field formation is displayed in Fig.~\ref{RFs0}. Around $M=64$, a rough structure of receptive field starts to emerge from the learning process, and
the structure becomes more apparent as more data is added. Meanwhile, the fraction of zero synapses decreases. Some of them become active, while some become inactive, refining
the developed receptive field or feature map. The belief about the stimulus image is continuously updated with more sensory inputs. Around $M=512$, a clear concept about digit $0$ is created by the unsupervised learning via combining likelihood and prior (see Eq.~(\ref{Pobs})).
Interestingly, a very small fraction ($4.2\%$) of zero synapses remain and serve as contour detectors. These zero synapses specify
the boundary between active and inactive regions in the feature map.

%%%%%%%%%%%%%%%%%%%%%%%%%%%%%%%%%%%%%%%%%%%%%%%%%%%%%%%%%%%%%%%%%%%%%%%
\begin{figure}
\centering
\subfigure[]{
     \includegraphics[bb=0 0 428 308,scale=0.57]{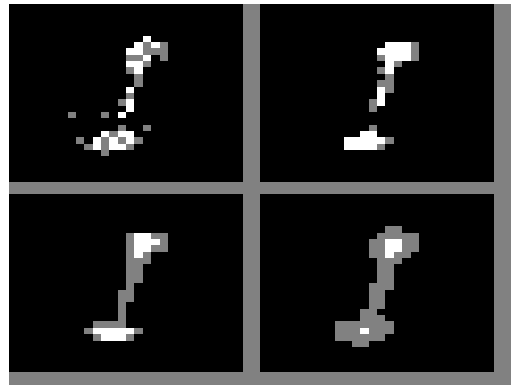}}
     \subfigure[]{\includegraphics[bb=0 0 326 243,scale=0.7]{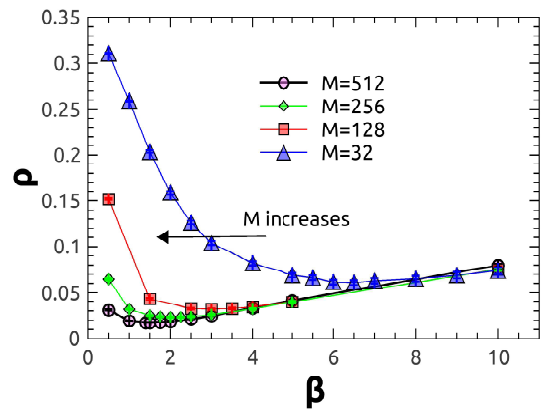}}
  \caption{
  (Color online) (a) Feature maps at different inverse-temperatures $\beta$. From left to right and from top to bottom, $\beta=0.5,2.0,5.0$
  and $10.0$. All feature maps are obtained when $M=512$. (b) The fraction of zero synapses versus the inverse-temperature at different values of $M$. Each marker in the plot is averaged over ten random selections of training images with equal number.
  }\label{Ebeta}
\end{figure}

%%%%%%%%%%%%%%%%%%%%%%%%%%%%%%%%%%%%%%%%%%

Next, we study the effect of the inverse-temperature $\beta$ on the receptive field formation. $\beta$ can be thought of as a scalar tuning the global contrast level of
the input image~\cite{Orban-2016}. By increasing $\beta$, one observes a qualitative change of the feature map (Fig.~\ref{Ebeta} (a)), from
an active-synapses-dominated phase (in the center of the feature map) at small $\beta$ to a zero-synapses-dominated phase at high $\beta$. Surprisingly, the zero-synapses-dominated phase 
still maintains the discriminative capability to distinguish different stimuli. Note that, at large $\beta$, the free energy ceases to be extensive, which can be seen from the last product of the second equality in Eq.~(\ref{Pobs}).
The qualitative change results from the competition between data constraints and biases introduced by zero synapses (Eq.~(\ref{Pobs})). The critical $\beta$ is determined by the value from which $\rho$ ceases to decrease and starts to increase. As observed in Fig.~\ref{Ebeta} (b), $\beta_c$ decreases as
$M$ increases.

\begin{figure}
\centering
     \includegraphics[bb=0 0 337 244,scale=0.7]{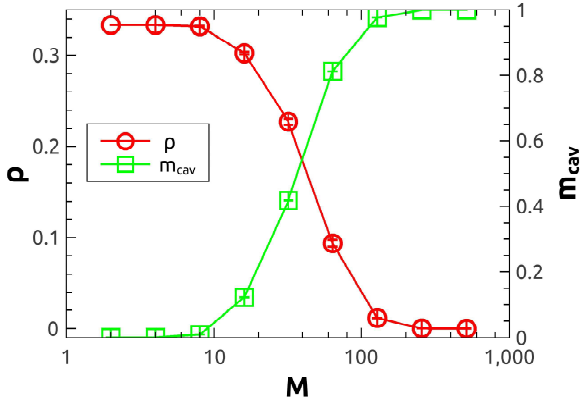}
  \caption{
  (Color online) The fraction $\rho$ of zero synapses (left axis) and the overall strength of cavity messages ($m_{{\rm cav}}$, right axis) as a function of the data size of neural activity.
  Each marker in the plot is the average over ten random selections of neural spike patterns with equal number.
  }\label{retina}
\end{figure}

Finally, we apply the computational framework to model retinal neural activity. We study a dataset composed of about $280\times10^3$ spike patterns of $160$ retinal ganglion cells.
The neural activity was measured during a natural-movie-stimuli experiment on the salamander retina (data courtesy of Michael J. Berry~\cite{Marre-2012}). The retina is 
an early visual system performing decorrelation computation of redundant visual inputs~\cite{Mark-2012}. The downstream brain areas may directly 
model the structure of population activity from the upstream area (such as retina), without any reference to external sensory inputs~\cite{Berry-2016}. Thus it is important to test our theory on this kind of unsupervised learning
of retinal neural activity. In Fig.~\ref{retina}, we observe similar behavior of 
the sparsity of synapses as found in learning a handwritten digits dataset. The learned feature map has a spontaneous symmetry breaking at some critical data size, where the sparsity of synapses
changes rapidly as well. After the spontaneous symmetry breaking, the feature map has two possible phases: synapses are either all-active or all-inactive, and the fraction of 
zero synapses becomes nearly zero. The hidden neuron in our model can be thought of as a unit in a downstream circuit along the ventral visual pathway, and the polarization of its receptive
field does not show any
intrinsic structures similar to those we already observed in learning a handwritten digits dataset. This may be because the retina circuit is at the bottom level of the visual hierarchy, while the concept of the visual input
can only be formed at the higher level of the cortical hierarchy~\cite{DiCarlo-2012}.
\section{Discussion}
%%%%%%%%%%%%%%%%%%%%%%%%%%%%%%%%%%%%%%%%%%%%%%%%%%%%%%%%%
In conclusion, we build a physics model of sparse unsupervised feature learning based on the one-bit RBM, and in this model, the sparseness of synaptic activity is
automatically learned from the noisy data. A rapid decrease of the number of zero synapses signals concept formation in the neural network, and the remaining zero synapses refine the learned concept 
by serving as contour detectors. In addition, zero synapses are sensitive to the contrast level of sensory inputs. 
These predictions may guide future neurobiological experiments.
In particular, the fact that the number of zero synapses acts as an indicator of concept formation is intimately related to the spontaneous symmetry breaking in the model.
These findings may also have implications on promising deep neuromorphic computation with discrete synapses~\cite{sparnet-2016}. 

It would be very interesting, yet challenging, to generalize the current 
framework to neural networks with multiple hidden neurons, and furthermore with hierarchical multi-layered architectures.

Previous studies showed that parallel retrieval of memory is possible by a random 
dilution of connections in random RBMs~\cite{Barra-2012,Barra-2014}, which may have connections to our current findings, in the sense that zero synapses offer
the possibility to simultaneously recall multiple patterns.
Furthermore, our findings on unsupervised learning with zero synapses are consistent with results reported in~\cite{Brunel-2004}, where supervised learning in a perceptron model of
cerebellar Purkinje cells was studied. An intuitive explanation is that, learning stretches the synaptic-weight distribution, pushing synapses towards their limit values (either $0$ in
the perceptron model~\cite{Brunel-2004} or $\pm1$ here)~\cite{brunel}. A recent work derived the paramagnetic-spin-glass transition line in a generalized RBM with spin and weight priors
interpolating between Gaussian and binary distributions~\cite{Barra-2017}, which may connect to our results of spontaneous symmetry breaking and concept formation in real data analysis.

%%%%%%%%%%%%%%%%%%%%%%%%%%%%%%%%%%%%%%%%%%%%%%%%%%%%%%%
\section*{Acknowledgments}

%\begin{acknowledgments}
I thank Taro Toyoizumi for his comments on silent synapses, Jack Raymond and James Humble for 
careful reading the manuscript, and Adriano Barra for drawing my attention to his previous works.
This research was supported by AMED under Grant Number JP15km0908001.
%\end{acknowledgments}
%%%%%%%%%%%%%%%%%%%%%%%%%%%%%%%%%%%%%%%%%%%%%%%%%%%%%%%%%%%%%%%

%\bibliography{ref}

\begin{thebibliography}{28}
\expandafter\ifx\csname natexlab\endcsname\relax\def\natexlab#1{#1}\fi
\expandafter\ifx\csname bibnamefont\endcsname\relax
  \def\bibnamefont#1{#1}\fi
\expandafter\ifx\csname bibfnamefont\endcsname\relax
  \def\bibfnamefont#1{#1}\fi
\expandafter\ifx\csname citenamefont\endcsname\relax
  \def\citenamefont#1{#1}\fi
\expandafter\ifx\csname url\endcsname\relax
  \def\url#1{\texttt{#1}}\fi
\expandafter\ifx\csname urlprefix\endcsname\relax\def\urlprefix{URL }\fi
\providecommand{\bibinfo}[2]{#2}
\providecommand{\eprint}[2][]{\url{#2}}

\bibitem[{\citenamefont{Lennie}(2003)}]{Len-2003}
\bibinfo{author}{\bibfnamefont{P.}~\bibnamefont{Lennie}},
  \bibinfo{journal}{Current Biology} \textbf{\bibinfo{volume}{13}},
  \bibinfo{pages}{493} (\bibinfo{year}{2003}).

\bibitem[{\citenamefont{Huang and Toyoizumi}(2016{\natexlab{a}})}]{Huang-2016a}
\bibinfo{author}{\bibfnamefont{H.}~\bibnamefont{Huang}} \bibnamefont{and}
  \bibinfo{author}{\bibfnamefont{T.}~\bibnamefont{Toyoizumi}},
  \bibinfo{journal}{Phys. Rev. E} \textbf{\bibinfo{volume}{93}},
  \bibinfo{pages}{062416} (\bibinfo{year}{2016}{\natexlab{a}}).

\bibitem[{\citenamefont{{Olshausen Bruno A.} and {Field David
  J.}}(1996)}]{SC-1996}
\bibinfo{author}{\bibnamefont{{Olshausen Bruno A.}}} \bibnamefont{and}
  \bibinfo{author}{\bibnamefont{{Field David J.}}}, \bibinfo{journal}{Nature}
  \textbf{\bibinfo{volume}{381}}, \bibinfo{pages}{607} (\bibinfo{year}{1996}).

\bibitem[{\citenamefont{Lee et~al.}(2008)\citenamefont{Lee, Ekanadham, and
  Ng}}]{Lee-2008}
\bibinfo{author}{\bibfnamefont{H.}~\bibnamefont{Lee}},
  \bibinfo{author}{\bibfnamefont{C.}~\bibnamefont{Ekanadham}},
  \bibnamefont{and} \bibinfo{author}{\bibfnamefont{A.~Y.} \bibnamefont{Ng}}, in
  \emph{\bibinfo{booktitle}{{Advances in Neural Information Processing Systems
  20}}}, edited by \bibinfo{editor}{\bibfnamefont{J.~C.} \bibnamefont{Platt}},
  \bibinfo{editor}{\bibfnamefont{D.}~\bibnamefont{Koller}},
  \bibinfo{editor}{\bibfnamefont{Y.}~\bibnamefont{Singer}}, \bibnamefont{and}
  \bibinfo{editor}{\bibfnamefont{S.~T.} \bibnamefont{Roweis}}
  (\bibinfo{publisher}{Curran Associates, Inc.}, \bibinfo{year}{2008}), pp.
  \bibinfo{pages}{873--880}.

\bibitem[{\citenamefont{Brunel et~al.}(2004)\citenamefont{Brunel, Hakim, Isope,
  Nadal, and Barbour}}]{Brunel-2004}
\bibinfo{author}{\bibfnamefont{N.}~\bibnamefont{Brunel}},
  \bibinfo{author}{\bibfnamefont{V.}~\bibnamefont{Hakim}},
  \bibinfo{author}{\bibfnamefont{P.}~\bibnamefont{Isope}},
  \bibinfo{author}{\bibfnamefont{J.-P.} \bibnamefont{Nadal}}, \bibnamefont{and}
  \bibinfo{author}{\bibfnamefont{B.}~\bibnamefont{Barbour}},
  \bibinfo{journal}{Neuron} \textbf{\bibinfo{volume}{43}}, \bibinfo{pages}{745}
  (\bibinfo{year}{2004}).

\bibitem[{\citenamefont{Barbour et~al.}(2007)\citenamefont{Barbour, Brunel,
  Hakim, and Nadal}}]{Barb-2007}
\bibinfo{author}{\bibfnamefont{B.}~\bibnamefont{Barbour}},
  \bibinfo{author}{\bibfnamefont{N.}~\bibnamefont{Brunel}},
  \bibinfo{author}{\bibfnamefont{V.}~\bibnamefont{Hakim}}, \bibnamefont{and}
  \bibinfo{author}{\bibfnamefont{J.-P.} \bibnamefont{Nadal}},
  \bibinfo{journal}{Trends in Neurosciences} \textbf{\bibinfo{volume}{30}},
  \bibinfo{pages}{622} (\bibinfo{year}{2007}).

\bibitem[{\citenamefont{Erhan et~al.}(2010)\citenamefont{Erhan, Bengio,
  Courville, Manzagol, Vincent, and Bengio}}]{Erhan-2010}
\bibinfo{author}{\bibfnamefont{D.}~\bibnamefont{Erhan}},
  \bibinfo{author}{\bibfnamefont{Y.}~\bibnamefont{Bengio}},
  \bibinfo{author}{\bibfnamefont{A.}~\bibnamefont{Courville}},
  \bibinfo{author}{\bibfnamefont{P.-A.} \bibnamefont{Manzagol}},
  \bibinfo{author}{\bibfnamefont{P.}~\bibnamefont{Vincent}}, \bibnamefont{and}
  \bibinfo{author}{\bibfnamefont{S.}~\bibnamefont{Bengio}},
  \bibinfo{journal}{J. Mach. Learn. Res.} \textbf{\bibinfo{volume}{11}},
  \bibinfo{pages}{625} (\bibinfo{year}{2010}).

\bibitem[{\citenamefont{Agliari et~al.}(2012)\citenamefont{Agliari, Barra,
  Galluzzi, Guerra, and Moauro}}]{Barra-2012}
\bibinfo{author}{\bibfnamefont{E.}~\bibnamefont{Agliari}},
  \bibinfo{author}{\bibfnamefont{A.}~\bibnamefont{Barra}},
  \bibinfo{author}{\bibfnamefont{A.}~\bibnamefont{Galluzzi}},
  \bibinfo{author}{\bibfnamefont{F.}~\bibnamefont{Guerra}}, \bibnamefont{and}
  \bibinfo{author}{\bibfnamefont{F.}~\bibnamefont{Moauro}},
  \bibinfo{journal}{Phys. Rev. Lett.} \textbf{\bibinfo{volume}{109}},
  \bibinfo{pages}{268101} (\bibinfo{year}{2012}).

\bibitem[{\citenamefont{Sollich et~al.}(2014)\citenamefont{Sollich, Tantari,
  Annibale, and Barra}}]{Barra-2014}
\bibinfo{author}{\bibfnamefont{P.}~\bibnamefont{Sollich}},
  \bibinfo{author}{\bibfnamefont{D.}~\bibnamefont{Tantari}},
  \bibinfo{author}{\bibfnamefont{A.}~\bibnamefont{Annibale}}, \bibnamefont{and}
  \bibinfo{author}{\bibfnamefont{A.}~\bibnamefont{Barra}},
  \bibinfo{journal}{Phys. Rev. Lett.} \textbf{\bibinfo{volume}{113}},
  \bibinfo{pages}{238106} (\bibinfo{year}{2014}).

\bibitem[{\citenamefont{Tubiana and Monasson}(2017)}]{Remi-2016}
\bibinfo{author}{\bibfnamefont{J.}~\bibnamefont{Tubiana}} \bibnamefont{and}
  \bibinfo{author}{\bibfnamefont{R.}~\bibnamefont{Monasson}},
  \bibinfo{journal}{Phys. Rev. Lett.} \textbf{\bibinfo{volume}{118}},
  \bibinfo{pages}{138301} (\bibinfo{year}{2017}).

\bibitem[{\citenamefont{Smolensky}(1986)}]{Smolensky-1986}
\bibinfo{author}{\bibfnamefont{P.}~\bibnamefont{Smolensky}}
  (\bibinfo{publisher}{MIT Press}, \bibinfo{address}{Cambridge, MA, USA},
  \bibinfo{year}{1986}), chap. \bibinfo{chapter}{Information Processing in
  Dynamical Systems: Foundations of Harmony Theory}, pp.
  \bibinfo{pages}{194--281}.

\bibitem[{\citenamefont{Huang and Toyoizumi}(2016{\natexlab{b}})}]{Huang-2016b}
\bibinfo{author}{\bibfnamefont{H.}~\bibnamefont{Huang}} \bibnamefont{and}
  \bibinfo{author}{\bibfnamefont{T.}~\bibnamefont{Toyoizumi}},
  \bibinfo{journal}{Phys. Rev. E} \textbf{\bibinfo{volume}{94}},
  \bibinfo{pages}{062310} (\bibinfo{year}{2016}{\natexlab{b}}).

\bibitem[{\citenamefont{Petersen et~al.}(1998)\citenamefont{Petersen, Malenka,
  Nicoll, and Hopfield}}]{Peter-1998}
\bibinfo{author}{\bibfnamefont{C.~C.~H.} \bibnamefont{Petersen}},
  \bibinfo{author}{\bibfnamefont{R.~C.} \bibnamefont{Malenka}},
  \bibinfo{author}{\bibfnamefont{R.~A.} \bibnamefont{Nicoll}},
  \bibnamefont{and} \bibinfo{author}{\bibfnamefont{J.~J.}
  \bibnamefont{Hopfield}}, \bibinfo{journal}{Proc. Nat. Acad. Sci.}
  \textbf{\bibinfo{volume}{95}}, \bibinfo{pages}{4732} (\bibinfo{year}{1998}).

\bibitem[{\citenamefont{O'Connor et~al.}(2005)\citenamefont{O'Connor,
  Wittenberg, and Wang}}]{Connor-2005}
\bibinfo{author}{\bibfnamefont{D.~H.} \bibnamefont{O'Connor}},
  \bibinfo{author}{\bibfnamefont{G.~M.} \bibnamefont{Wittenberg}},
  \bibnamefont{and} \bibinfo{author}{\bibfnamefont{S.~S.-H.}
  \bibnamefont{Wang}}, \bibinfo{journal}{Proc. Nat. Acad. Sci.}
  \textbf{\bibinfo{volume}{102}}, \bibinfo{pages}{9679} (\bibinfo{year}{2005}).

\bibitem[{\citenamefont{Huang}(2017)}]{Huang-2017}
\bibinfo{author}{\bibfnamefont{H.}~\bibnamefont{Huang}},
  \bibinfo{journal}{Journal of Statistical Mechanics: Theory and Experiment}
  \textbf{\bibinfo{volume}{2017}}, \bibinfo{pages}{053302}
  (\bibinfo{year}{2017}).

\bibitem[{\citenamefont{Lecun et~al.}(1998)\citenamefont{Lecun, Bottou, Bengio,
  and Haffner}}]{Lecun-1998}
\bibinfo{author}{\bibfnamefont{Y.}~\bibnamefont{Lecun}},
  \bibinfo{author}{\bibfnamefont{L.}~\bibnamefont{Bottou}},
  \bibinfo{author}{\bibfnamefont{Y.}~\bibnamefont{Bengio}}, \bibnamefont{and}
  \bibinfo{author}{\bibfnamefont{P.}~\bibnamefont{Haffner}},
  \bibinfo{journal}{Proceedings of the IEEE} \textbf{\bibinfo{volume}{86}},
  \bibinfo{pages}{2278} (\bibinfo{year}{1998}).

\bibitem[{\citenamefont{MacKay}(1992)}]{MacKay-1992}
\bibinfo{author}{\bibfnamefont{D.~J.~C.} \bibnamefont{MacKay}},
  \bibinfo{journal}{Neural Comput.} \textbf{\bibinfo{volume}{4}},
  \bibinfo{pages}{448} (\bibinfo{year}{1992}).

\bibitem[{\citenamefont{Nishimori}(2001)}]{Nishimori-2001}
\bibinfo{author}{\bibfnamefont{H.}~\bibnamefont{Nishimori}},
  \emph{\bibinfo{title}{Statistical Physics of Spin Glasses and Information
  Processing: An Introduction}} (\bibinfo{publisher}{Oxford University Press},
  \bibinfo{address}{Oxford}, \bibinfo{year}{2001}).

\bibitem[{\citenamefont{M\'ezard and Montanari}(2009)}]{MM-2009}
\bibinfo{author}{\bibfnamefont{M.}~\bibnamefont{M\'ezard}} \bibnamefont{and}
  \bibinfo{author}{\bibfnamefont{A.}~\bibnamefont{Montanari}},
  \emph{\bibinfo{title}{Information, Physics, and Computation}}
  (\bibinfo{publisher}{Oxford University Press}, \bibinfo{address}{Oxford},
  \bibinfo{year}{2009}).

\bibitem[{\citenamefont{Huang and Toyoizumi}(2015)}]{Huang-2015b}
\bibinfo{author}{\bibfnamefont{H.}~\bibnamefont{Huang}} \bibnamefont{and}
  \bibinfo{author}{\bibfnamefont{T.}~\bibnamefont{Toyoizumi}},
  \bibinfo{journal}{Phys. Rev. E} \textbf{\bibinfo{volume}{91}},
  \bibinfo{pages}{050101} (\bibinfo{year}{2015}).

\bibitem[{\citenamefont{Orban et~al.}(2016)\citenamefont{Orban, Berkes, Fiser,
  and Lengyel}}]{Orban-2016}
\bibinfo{author}{\bibfnamefont{G.}~\bibnamefont{Orban}},
  \bibinfo{author}{\bibfnamefont{P.}~\bibnamefont{Berkes}},
  \bibinfo{author}{\bibfnamefont{J.}~\bibnamefont{Fiser}}, \bibnamefont{and}
  \bibinfo{author}{\bibfnamefont{M.}~\bibnamefont{Lengyel}},
  \bibinfo{journal}{Neuron} \textbf{\bibinfo{volume}{92}}, \bibinfo{pages}{530}
  (\bibinfo{year}{2016}).

\bibitem[{\citenamefont{Marre et~al.}(2012)\citenamefont{Marre, Amodei,
  Deshmukh, Sadeghi, Soo, Holy, and Berry}}]{Marre-2012}
\bibinfo{author}{\bibfnamefont{O.}~\bibnamefont{Marre}},
  \bibinfo{author}{\bibfnamefont{D.}~\bibnamefont{Amodei}},
  \bibinfo{author}{\bibfnamefont{N.}~\bibnamefont{Deshmukh}},
  \bibinfo{author}{\bibfnamefont{K.}~\bibnamefont{Sadeghi}},
  \bibinfo{author}{\bibfnamefont{F.}~\bibnamefont{Soo}},
  \bibinfo{author}{\bibfnamefont{T.~E.} \bibnamefont{Holy}}, \bibnamefont{and}
  \bibinfo{author}{\bibfnamefont{M.~J.} \bibnamefont{Berry}},
  \bibinfo{journal}{J. Neurosci.} \textbf{\bibinfo{volume}{32}},
  \bibinfo{pages}{14859} (\bibinfo{year}{2012}).

\bibitem[{\citenamefont{{Pitkow Xaq} and {Meister Markus}}(2012)}]{Mark-2012}
\bibinfo{author}{\bibnamefont{{Pitkow Xaq}}} \bibnamefont{and}
  \bibinfo{author}{\bibnamefont{{Meister Markus}}}, \bibinfo{journal}{Nat
  Neurosci} \textbf{\bibinfo{volume}{15}}, \bibinfo{pages}{628}
  (\bibinfo{year}{2012}).

\bibitem[{\citenamefont{{Loback} et~al.}(2016)\citenamefont{{Loback},
  {Prentice}, {Ioffe}, and {Berry}}}]{Berry-2016}
\bibinfo{author}{\bibfnamefont{A.~R.} \bibnamefont{{Loback}}},
  \bibinfo{author}{\bibfnamefont{J.~S.} \bibnamefont{{Prentice}}},
  \bibinfo{author}{\bibfnamefont{M.~L.} \bibnamefont{{Ioffe}}},
  \bibnamefont{and} \bibinfo{author}{\bibfnamefont{M.~J.}
  \bibnamefont{{Berry}}, \bibfnamefont{II}}, \bibinfo{journal}{ArXiv e-prints
  1610.06886}  (\bibinfo{year}{2016}).

\bibitem[{\citenamefont{DiCarlo et~al.}(2012)\citenamefont{DiCarlo, Zoccolan,
  and Rust}}]{DiCarlo-2012}
\bibinfo{author}{\bibfnamefont{J.~J.} \bibnamefont{DiCarlo}},
  \bibinfo{author}{\bibfnamefont{D.}~\bibnamefont{Zoccolan}}, \bibnamefont{and}
  \bibinfo{author}{\bibfnamefont{N.~C.} \bibnamefont{Rust}},
  \bibinfo{journal}{Neuron} \textbf{\bibinfo{volume}{73}}, \bibinfo{pages}{415}
  (\bibinfo{year}{2012}).

\bibitem[{\citenamefont{{Ardakani} et~al.}(2016)\citenamefont{{Ardakani},
  {Condo}, and {Gross}}}]{sparnet-2016}
\bibinfo{author}{\bibfnamefont{A.}~\bibnamefont{{Ardakani}}},
  \bibinfo{author}{\bibfnamefont{C.}~\bibnamefont{{Condo}}}, \bibnamefont{and}
  \bibinfo{author}{\bibfnamefont{W.~J.} \bibnamefont{{Gross}}},
  \bibinfo{journal}{ArXiv e-prints 1611.01427}  (\bibinfo{year}{2016}).

\bibitem[{bru()}]{brunel}
\bibinfo{note}{Nicolas Brunel, personal communication}.

\bibitem[{\citenamefont{{Barra} et~al.}(2017)\citenamefont{{Barra}, {Genovese},
  {Sollich}, and {Tantari}}}]{Barra-2017}
\bibinfo{author}{\bibfnamefont{A.}~\bibnamefont{{Barra}}},
  \bibinfo{author}{\bibfnamefont{G.}~\bibnamefont{{Genovese}}},
  \bibinfo{author}{\bibfnamefont{P.}~\bibnamefont{{Sollich}}},
  \bibnamefont{and}
  \bibinfo{author}{\bibfnamefont{D.}~\bibnamefont{{Tantari}}},
  \bibinfo{journal}{ArXiv e-prints: 1702.05882}  (\bibinfo{year}{2017}).

\end{thebibliography}

%%%%%%%%%%%%%%%%%%%%%%%%%%%%%%%%%%%%%%%%%%%%%%%%%%%%%%%%%%%%%%%%%%%%%

\end{document}